\documentclass{article}
\usepackage{amsmath}
\usepackage{authblk}
\usepackage{cite}
\usepackage{graphicx}

\newtheorem{lemma}{Lemma}

\title{Tensor Evolution: A Framework for Fast Evaluation of Tensor Computations using Recurrences}

\author[1]{Javed Absar}
\author[2]{Samarth Narang }
\author[3]{Muthu Baskaran }
\affil[1]{Qualcomm Technologies International}
\affil[2]{Qualcomm Technologies, Inc.}
\affil[3]{Qualcomm Technologies, Inc.}

\begin{document}
\maketitle

\begin{abstract}
This paper introduces a new mathematical framework for analysis and optimization of tensor
expressions within an enclosing loop. Tensors are multi-dimensional arrays of values.
They are common in high performance computing (HPC)  and machine learning
\cite{TensorComprehension} domains. Our framework extends Scalar Evolution
\cite{scev_j} -- an important optimization pass implemented in both LLVM and GCC -- to tensors.
Scalar Evolution (SCEV) relies on the theory of `Chain of Recurrences' \cite{Zima} for its
mathematical underpinnings. We use the same theory for \textbf{Tensor Evolution} (TeV).
While some concepts from SCEV map easily to TeV -- e.g. element-wise operations;
tensors introduce new operations \cite{Attention, mlir,TVM,Glow} such as concatenation,
slicing, broadcast, reduction, and reshape which have no equivalent in scalars and SCEV.
Not all computations are amenable to TeV analysis but it can play a part in the optimization
and analysis parts of ML and HPC compilers. Also, for many mathematical/compiler ideas,
applications may go beyond what was initially envisioned, once others build on it and
take it further. We hope for a similar trajectory for the tensor-evolution concept.
\end{abstract}

\section{Introduction}
In this work, we introduce \textbf{Tensor Evolution} (TeV), a framework designed for analysis and optimization
 of tensor computations within loops. TeV provides a mathematical framework and mechanism to trace tensor
  values as they evolve over iterations of the enclosing loops. This approach enables optimizations that
   simplify loop-carried tensor computations to reduce computational overhead.

As machine learning (ML) models grow in complexity and scale, the demand for efficient computation has
spurred the development of specialized ML compilers, such as MLIR\cite{mlir}, Glow\cite{Glow}, and
TVM\cite{TVM}. These compilers convert computation in frameworks such as PyTorch and TensorFlow into
IR (intermediate representation) graphs where nodes represent operations
(e.g., matrix multiplications, additions, etc.), and edges represent tensors.
While traditional compiler optimizations, such as Constant Propagation, Loop Strength Reduction,
and Loop Invariant Code Motion have been somewhat adapted to these graph compilers,
more advanced transformations for tensor operations remain unexplored. Notably,
scalar evolution which is widely successful in traditional compilers such as LLVM for
analyzing scalar variables in loops lacks a counterpart in tensor operations within deep learning models or HPC. 

The paper is organized as follows: To explain TeV, we first introduce the basic concepts of recurrences
 and SCEV. Then, we present a formal definition of TeV. We outline a set of lemmas (re-write rules)
  that simplify TeV expressions. Next, we provide a worked-out example. Finally, we survey related
   work and conclude the paper.

Please note that part of this work was previously presented at the LLVM Developer
Summit \cite{tev_j} in the form of `Technical Talk`.

\section{Chain of Recurrences and Scalar Evolution}
Consider the function $S(n) = 1+2+3...+n$. We know that $S(n)=n(n+1)/2$, and this closed form can help
 calculate $S$ at any point, bypassing the need for a long chain of sequential additions.
  `Chain of Recurrences' (CR) is essentially a more complex algebraic generalization of this
concept \cite{ZimaRealPaper}.CRs are a compact representation of a series of computations. 
They provide a set of lemmas to simplify CR expressions and offer a closed-form function to evaluate
 these representations at any fixed point. Scalar Evolution (SCEV), implemented in LLVM and GCC,
  incorporates a simplified form of CR. To understand CR and SCEV, let us consider an example.

\begin{verbatim}
    foo(...) { 
      int f = k0;
      for (int i = 0; i < n; ++i) {
         ... 
         f = f + k1;
      }
      ... = f // loop exit  value of `f`
\end{verbatim}

Here the value of $f$ can be expressed as a `basic recurrence'($BR$) $f = \{k_0, +, k_1\}_i$. 
In this $BR$ notation, $f$ has initial value $k_0$ and in each iteration of the loop with 
induction variable $i$, $f$ increases by $k_1$. In compiler terminology, $i$ is a basic 
induction variable, while $f$ is a derived induction variable. To help appreciate the algebraic
 power of recurrences, we show a few lemmas on basic recurrence.
    

\section{Basic Recurrence Lemmas}
Let $f = \{a, \odot, b\}$ represent a basic recurrence, where $\odot$ denotes an operator (e.g., addition or multiplication), and let $c$ be a constant. The following lemmas express  properties of basic recurrences:
    
    \begin{lemma}[Addition of a Constant]
    For a constant $c$ added to a basic recurrence:
    \[
    c + \{a, +, b\} \implies \{c + a, +, b\}.
    \]
    \end{lemma}
    
    \begin{lemma}[Multiplication by a Constant]
    For a constant $c$ multiplied with a basic recurrence:
    \[
    c \cdot \{a, +, b\} \implies \{c \cdot a, +, c \cdot b\}.
    \]
    \end{lemma}
    
    \begin{lemma}[Exponentiation by a Constant]
    For a constant $c$ raised to the power of a basic recurrence:
    \[
    c^{\{a, +, b\}} \implies \{c^a, \cdot, c^b\}.
    \]
    \end{lemma}
    
    \section*{Combining Basic Recurrences}
    
    Let $f = \{a, \odot, b\}$ and $g = \{c, \odot, d\}$ represent two basic recurrences. It can be proven that:
    
    \begin{lemma}[Addition of Two Recurrences]
    For the addition of two basic recurrences:
    \[
    \{a, +, b\} + \{c, +, d\} \implies \{a + c, +, b + d\}.
    \]
    \end{lemma}
    
    \begin{lemma}[Multiplication of Two Recurrences]
    For the multiplication of two basic recurrences:
    \[
    \{a, +, b\} \cdot \{c, +, d\} \implies \{ac, +, \{a, \odot, b\}d + \{c, \odot, d\}b + bd\}.
    \]
    \end{lemma}
    
    \section*{Chain of Recurrences}    
    A chain of recurrences (CR) is defined recursively as a function over $N$ with
     constants $\{c_0, c_1, \dots, c_{k-1}\}$, a function $f_k$, and an operator $\odot$ (either $+$ or $\cdot$). 
    
    \begin{lemma}[Chain of Recurrences]
    A chain of recurrences $\phi(i)$ is defined as:
    \[
    \phi(i) = \{c_0, \odot, \{c_1, \odot, c_2, \dots, \odot, f_k\}\}(i).
    \]
    \end{lemma}
    
    In practice, the nested braces are omitted, and the chain of recurrences
    simply written out as (LLVM SCEV prints out in same format):
    \[
    \phi = \{c_0, \odot, c_1, \odot, c_2, \dots, \odot, f_k\}.
    \]
    
    Consider the CR $\phi = \{7, +, 6, +, 10, +, 6\}$. This CR can be simplified to the closed expression:
    \[
    \phi(i) = i^3 + 2i^2 + 3i + 7.
    \]

This closed expression can be derived by applying the recurrence lemmas outlined
above and is described in detail in \cite{ZimaRealPaper, Zima, OlafRecurrence}.
The application of chains of recurrences in LLVM, including their usage for 
induction variables, is discussed in \cite{scev_j}.
    
    \section{Tensor Evolution}
    Both Scalar Evolution and Tensor Evolution are founded on the principle of `compact representation',
     which enables computational reducibility \cite{Wolfram}—the ability to compute the value at a distant
      iteration point directly, without sequentially processing each iteration of the loop.
    
    \subsection{Code Example}
    Consider the following code -    
    \begin{verbatim}
        def forward(a: torch.Tensor, x: torch.Tensor) -> torch.Tensor:
            for _ in range(15):
                x = a + x 
            return x
    \end{verbatim}

    By analyzing the code and understanding the loop's behavior, we observe that the
     return (final) value is  $a+15x$, where $x$ is the initial input tensor value.
      A tensor can be viewed as a collection of scalar values. Thus, we can see the 
      connection between what we refer to as \textit{Tensor Evolution} and the concepts discussed
       in the previous section—\textit{Scalar Evolution} and \textit{Chain of Recurrences}.

    \subsection{Basic Formulation}    
    \begin{figure}[h!]
    \centering
    \includegraphics[width=0.75\linewidth]{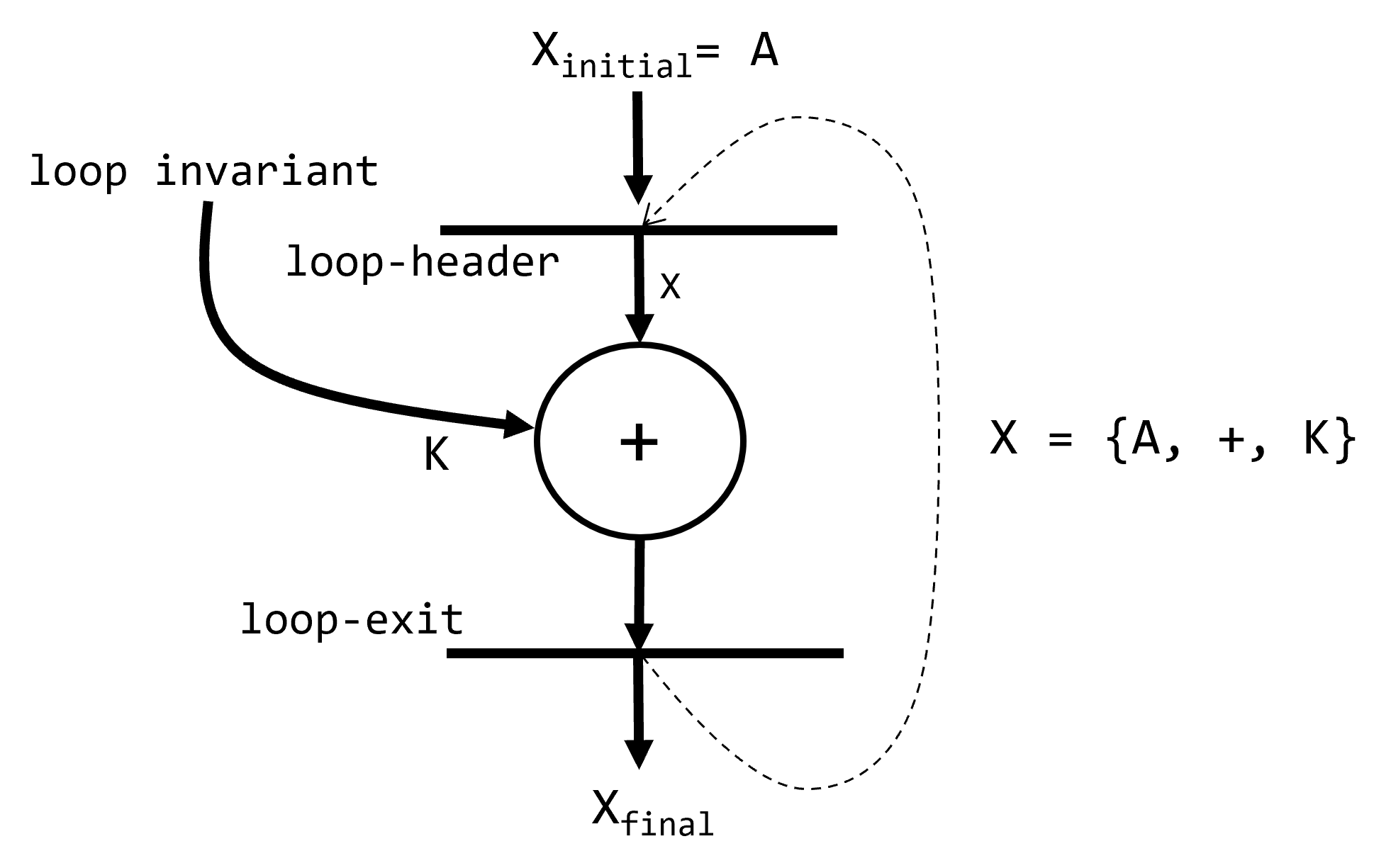}
    \caption{Basic Tensor Evolution}
    \label{fig:basic_TEV}
    \end{figure}

Given a constant or loop-invariant tensor $T_c$,  a function $\tau_1$ over the natural
numbers $N$ that produces a tensor of same shape as $T_c$, and an element-wise operator
$\odot$, then the basic evolution of the tensor value represented by tuple
$\tau=\{T_c, \odot, \tau_1\}$ is defined as a function $\tau(i)$ over $N$
(see Fig. \ref{fig:basic_TEV}):
    
\begin{equation}
\{T_c, \odot, \tau_1\}(i) = T_c \odot \tau_1(1) \odot \tau_1(2) ... \odot \tau_1(i)
\end{equation}
    
    For instance, given a zeros tensor $T_0$, a ones tensor $T_1$, and the operator `+'
     representing tensor addition, the tensor evolution expression $\tau=\{T_0, +, T_1\}$
      produces the sequence of tensors $T_0$, $T_1$, ..., $T_k$ for subsequent values
       of $\tau(i)$. Note in this case, the function $\tau_1(i) = T_i, \forall i \in N$.
    
    Next, given loop-invariant tensors $T_{c_0}, T_{c_1}, T_{c_2}, ...,T_{c_{i-1}}$,
     a function $\tau_k$ defined over $N$, and operators $\odot_{1}, \odot_{2}, ..., \odot_k$,
      a chain of evolution of the tensor value is represented by the tuple 
    
    \begin{equation}
    \tau = \{T_{c_0}, \odot_{1},T_{c_1}, \odot_{2}, ..., \odot_{k}, \tau_{k} \}    
    \end{equation}
    
    which is defined recursively as 
    \begin{equation}
    \tau(i) = \{T_{c_0}, \odot_{1}, \{T_{c_1}, \odot_{2}, ..., \odot_{k}, \tau_{k} \} \}(i)
    \end{equation}
    
    The operators $\odot_{i}$ could all be the same (e.g., $+$ or $*$) or different. Additionally,
     if we drop the braces and understand that the application of evolution proceeds 
     from right to left, the chain of evolution can be rewritten simply as: 
    
    \begin{equation}
    \tau(i) = \{T_{c_0}, \odot_{1}, T_{c_1}, \odot_{2}, ..., \odot_{k}, \tau_{k} \}(i)
    \end{equation}
    
    \section{Tensor Evolution Lemmas}
    Next, we develop a set of lemmas for Tensor Evolution. These lemmas are algebraic 
    properties (re-write rules) designed to simplify computations involving tensor evolution expressions.

    \begin{lemma}{Add or multiply TeV by a loop invariant tensor}
    \label{lem::evol_add_basic}
    \newline Let $\{A,\odot,\tau\}$ be a tensor evolution expression where $\odot$ is either element wise add, subtract or multiply (see Fig. \ref{lem::evol_add_basic}).
    And, let $K$ be a loop invariant tensor. Then, 
    \[
    K+\{A, +, \tau \} \Rightarrow \{K+A, +, \tau \}
    \]
    \[
    K *\{A,+,\tau\} \Rightarrow \{K*A,+,K*\tau \}
    \]
    \end{lemma}

    \begin{lemma}{Reshape of tensor evolution expression}
    \label{lem::Reshape}
    \newline Let $\{A,\odot,\tau\}$ be a tensor evolution expression where $\odot$ is an element-wise operator.
    Let $\mathcal{R}(A)$ represent the reshaping of the tensor 
    A, which includes the permutation of dimensions and the ordinary transpose.
    Then,
    \[
     \mathcal{R}(\{A, \odot, \tau \}) 
     \Rightarrow
     \{\mathcal{R}(A), \odot, \mathcal{R}(\tau)\}
    \]
    \end{lemma}
    

    \begin{lemma}{Slice of tensor evolution expression}
    \label{lem::Slice}
    \newline Let $\{A,\odot,\tau\}$ be a tensor evolution expression where $\odot$ is an element-wise operator.
    Let $\mathcal{S}(A)$ represent slicing of the tensor, where slicing selects a subset of elements to form the resulting tensor.
    Then,
    \[
     \mathcal{S}(\{A, \odot, \tau \}) 
     \Rightarrow 
     \{\mathcal{S}(A), \odot, \mathcal{S}(\tau)\}
    \]
    \end{lemma}
    
    
    \begin{lemma}{Concatenation of TeV expressions}
    \label{lem::Concat}
    \newline Let $\{A,\odot,\tau_1\}$ and $\{B,\odot,\tau_2\}$ be two tensor evolution expressions
    where $\odot$ is an element-wise operator.
    Let $\mathcal{C}(T_1,T_2)$ represent concatenation of two tensors.
    Then,
    \[
     \mathcal{C}(\{A,\odot,\tau_1\}, \{B,\odot,\tau_2\})
     \Rightarrow 
     \{\mathcal{C}(A,B),\odot,\mathcal{C}(\tau_1,\tau_2)\} 
    \]
    \end{lemma}
    
    \begin{figure}[h!]
    \centering
      \includegraphics[width=0.75\linewidth]{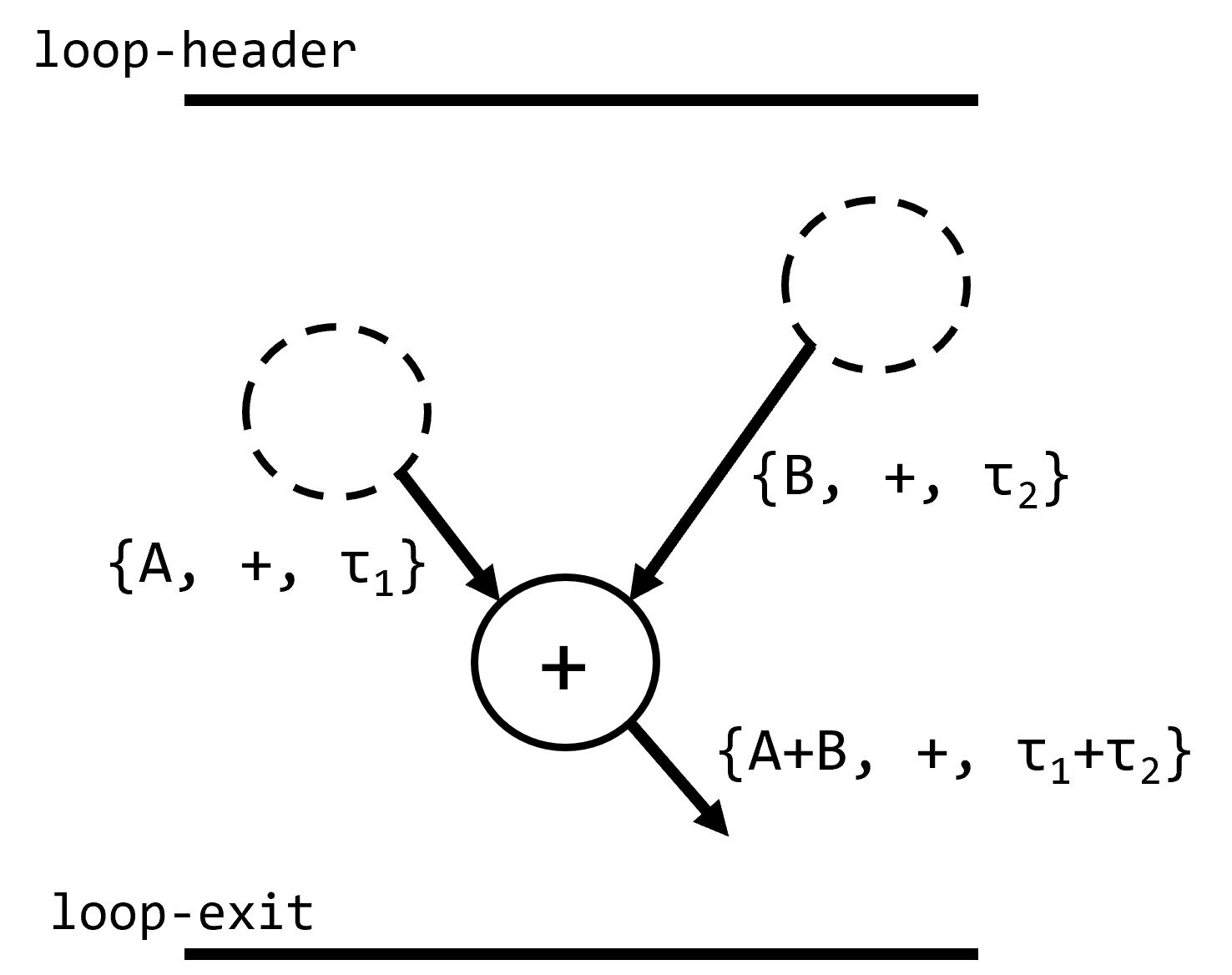}
      \caption{Adding two TeV expressions}
      \label{fig::evol_add_chain}
    \end{figure}

    \begin{lemma}{Adding two TeV Expressions}
    \label{lem::evol_add_chain}
    \newline Let $\{A,\odot,\tau_1\}$ and $\{B,\odot,\tau_2\}$ be two tensor evolution expressions (see Fig. \ref{fig::evol_add_chain}). Then, 
    \[
    \{A, +, \tau_1 \} + \{B, +, \tau_2\} \Rightarrow \{A+B, +, \tau_1 + \tau_2\}
    \]
    \end{lemma}
    
    
    \begin{lemma}{Multiply two TeV Expressions}
    \label{lem::evol_mul_chain}
    \newline Let $\{A,\odot,\tau_1\}$ and $\{B,\odot,\tau_2\}$ be two tensor evolution expressions. Then, 
    \[
    \{A, +, \tau_1 \} * \{B, +, \tau_2\} \Rightarrow \big\{AB, +, \tau_1\{B, +, \tau_2\} + \tau_2\{A, +, \tau_1 \} + \tau_1\tau_2\big\}
    \]
    \end{lemma}
    
   
    \begin{lemma}{TeV input to a loop variant tensor}
    \label{lem::evol_mem_chain}
    \newline Let $X$ be a loop variant tensor with an initial value of $A$. At each iteration,
     its value evolves as $A + \{B,\odot_2,\tau_2\}$, meaning the TeV $\{B,\odot_2,\tau_2\}$ is 
     added to the current value of $X$. This is represented as 
    $\{B, +, \tau_2\} \xrightarrow[+]{} A$. Then,
    \newline
    \[
    \{B, +, \tau_2\} \xrightarrow[+]{} A \Rightarrow \{A, +, B, +, \tau\}
    \]
    In other words, this results in a chain of evolution of depth two.
    \end{lemma}
    
    There are additional lemmas based on variations in composition and operators beyond those
     discussed above. However, the lemmas presented so far are already sufficient to address
      a range of interesting problems.
    \begin{table}[h!]
    \centering
    \begin{tabular}{|c|c|c|}
    \hline
    \textbf{Operator} & \textbf{Tensor Expression} & \textbf{Rewrite-rule} \\ \hline
    reshape & $\mathcal{R}(\{A, +, \tau\})$ & $\{\mathcal{R}(A), +, \mathcal{R}(\tau)\}$ \\ 
            & $\mathcal{R}(\{A, \ast, \tau\})$ & $\{\mathcal{R}(A), \ast, \mathcal{R}(\tau)\}$ \\ \hline
    slice & $\mathcal{S}(\{A, +, \tau\})$ & $\{\mathcal{S}(A), +, \mathcal{S}(\tau)\}$ \\ 
          & $\mathcal{S}(\{A, \ast, \tau\})$ & $\{\mathcal{S}(A), \ast, \mathcal{S}(\tau)\}$ \\ \hline
    broadcast & $\mathcal{B}(\{A, +, \tau\})$ & $\{\mathcal{B}(A), +, \mathcal{B}(\tau)\}$ \\ 
              & $\mathcal{B}(\{A, \ast, \tau\})$ & $\{\mathcal{B}(A), \ast, \mathcal{B}(\tau)\}$ \\ \hline
    concat & $\mathcal{C}(\{\{A, +, \tau_1\}, \{B, \odot, \tau_2\}\})$ & $\{\mathcal{C}(A, B), +, \mathcal{C}(\tau_1, \tau_2)\}$ \\ 
           & $\mathcal{C}(\{\{A, \ast, \tau_1\}, \{B, \odot, \tau_2\}\})$ & $\{\mathcal{C}(A, B), \ast, \mathcal{C}(\tau_1, \tau_2)\}$ \\ \hline
    log & $\log(\{A, \ast, \tau\})$ & $\{\log(A), +, \log(\tau)\}$ \\ \hline
    exponent & $\exp(\{A, +, \tau\})$ & $\{\exp(A), \ast, \exp(\tau)\}$ \\ \hline
    add const & $K + \{A, +, \tau\}$ & $\{K + A, +, \tau\}$ \\ \hline
    multiply by const & $K \ast \{A, +, \tau\}$ & $\{K \ast A, +, K \ast \tau\}$ \\ \hline
    add two TeV & $\{A, +, \tau_1\} + \{B, +, \tau_2\}$ & $\{A + B, +, \tau_1 + \tau_2\}$ \\ \hline
    inject TeV into TeV & $\{B, +, \tau_2\} \to A$ & $\{A, +, B, +, \tau\}$ \\ \hline
    \end{tabular}
    \caption{Useful Lemmas (re-write Rules) for Tensor Evolution Evaluation.}
    \label{table:tev}
    \end{table}
    
    \section{Application of Tensor Evolution}
    
\begin{figure}[h!]
    \centering
      \includegraphics[width=\linewidth]{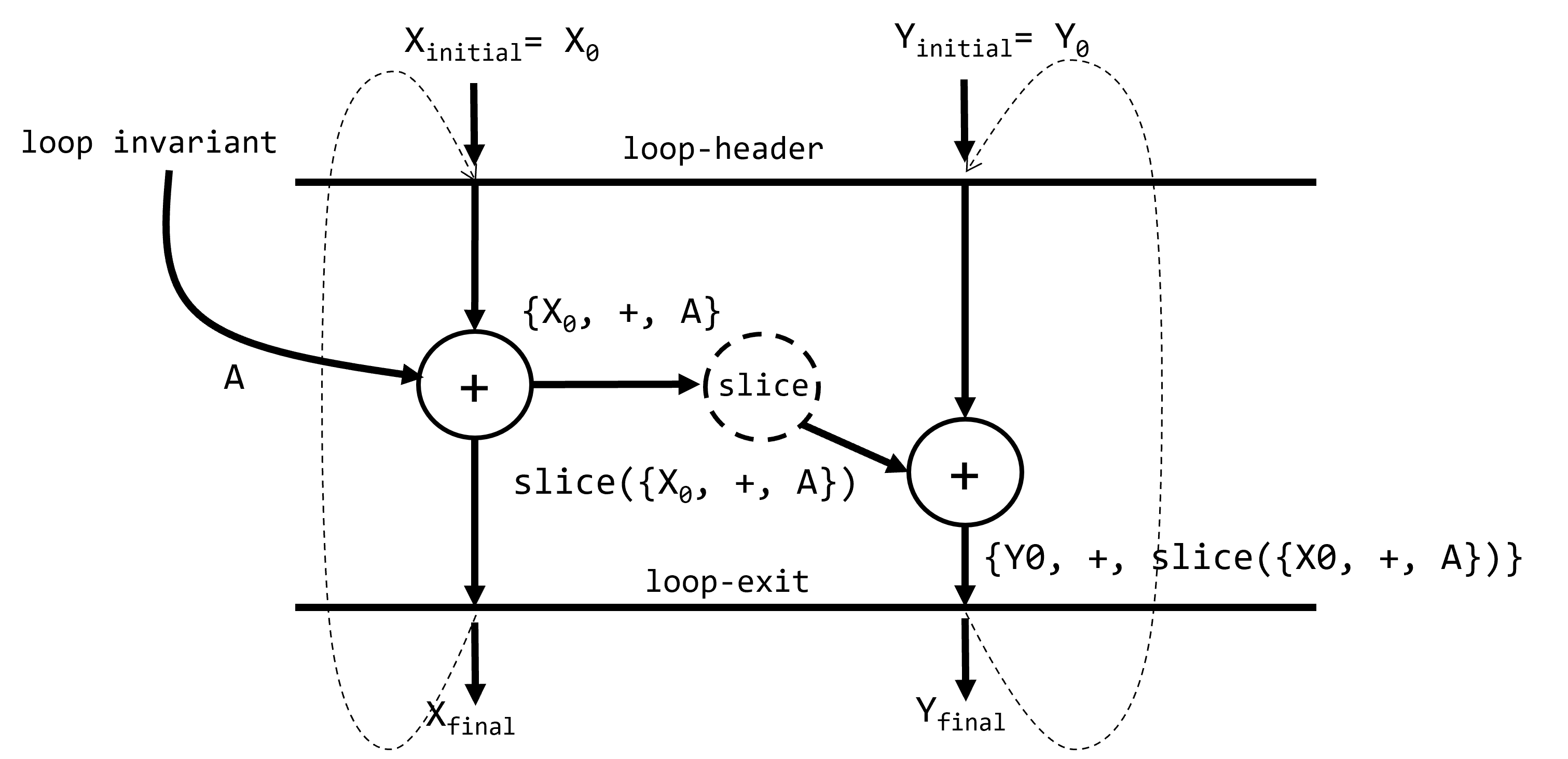}
      \caption{Calculate tensor evolution expressions of given code.}
      \label{fig:TEV_Code_Example}
    \end{figure}

    Consider the following code which operates on tensors -
    \begin{verbatim}
        def forward(a: torch.Tensor, x: torch.Tensor) -> torch.Tensor:
          y = torch.zeros_like(x)
          for _ in range(15):
            x = x + a
            z = x[1,:]
            y = y + z
          return y
    \end{verbatim}
    The interpretation of how the tensor values are evolving across iterations of the loop is shown in \ref{fig:TEV_Code_Example}. 
    
    Now, using the re-write rules of Tensor Evolution that were laid out before, we can construct and simplify the TeV expressions.
    
    \begin{equation}
    \begin{cases}
    Y_k = \left\{  Y_0, +, \mathcal{S}( \left\{ X_0, +, A\right\})    \right\}_k \\
    Y_k= \left\{  Y_0, +,  \left\{\mathcal{S}(X_0), +, \mathcal{S}(A)\right\}    \right\}_k \\
    Y_k= \left\{  Y_0, +,  \mathcal{S}(X_0), +, \mathcal{S}(A) \right\}_k \\
    Y_k= Y_0 + k * \mathcal{S}(X_0) + k*(k+1)/2 *  \mathcal{S}(A)    
    \end{cases}
    \end{equation}
    
    Based on the above simplifications (rewrites) the same code can be re-written now as 
    \begin{verbatim}
    def forward(self, a: torch.Tensor, x:torch.Tensor)->torch.Tensor: 
      y = torch.zeros_like(x)
      return y + 15*x[1,:] + 15*(15+1)/2*a[1,:]
    \end{verbatim}
    
    \section{Background Art and Related Work}
    Tensor Comprehensions (TC) \cite{TensorComprehension} is another representation for
     computations on tensors, based on generalized Einstein notation. For example, let $X$ and $Y$
      be two input tensors, and $Z$ the output tensor.  The TC statement $Z(r) += X(r,c) * Y(c)$ 
      introduces two index variables, $r$ and $c$. Their ranges are inferred from their indexing 
      of tensors $X$ and $Y$. Since $c$ appears only on the right-hand side, the computation reduces
       over $c$ when storing values into $Z$. TC provides a compact representation of computations. 
       However, it does not address the \textit{computational reducibility} aspect of the problem.
        In contrast, SCEV and TeV include algebraic rewrite rules that simplify computations. 
        While not all computations are reducible, areas of reducibility often exist in general code,
         and both TeV and SCEV are specifically designed to identify and simplify these reducible computations.
    
    As mentioned earlier, Tensor Evolution is based on the concept of Chain of Recurrences (CR).
     The paper by \cite{Zima} on CR develops rewrite rules for long chains and complex operations,
      including trigonometric functions and exponents of exponent operations. Compilers such as LLVM 
      and GCC implement CR as Scalar Evolution (SCEV), but only for simple operations like addition
       and subtraction, and limited to two levels of recurrence. While extending beyond this is 
       mathematically possible, it is not practical for general-purpose compilers and has limited
        usability in real-world applications.
     
    Many Domain-Specific Languages (DSLs) and compilers have been developed for machine 
    learning (ML) and high-performance computing (HPC). DSLs leverage domain-specific constructs
     to capture parallelism, locality, and other application-specific characteristics. 
     An example is Halide \cite{HALIDE}, a DSL for image processing. Halide defines its inputs
      as images over an infinite range, whereas Tensor Comprehensions (TC) sets fixed sizes for
       each dimension using range inference. In ML applications, most computations are performed 
       on fixed-size tensors with higher temporal locality than images. Additionally, TC is less
        verbose than Halide, as Halide carries the syntactic burden of pre-declaring stage names 
        and free variables.
    
    The polyhedral framework is another abstraction used for the analysis and transformation of loop nests.
     One of the first high-level language polyhedral compilers was the R-Stream 3.0 Compiler \cite{Lethin2008RStream3C}
      and several tools and libraries have been developed to harness its benefits, such as GCC (Graphite) and LLVM
       (Polly). Polyhedral techniques have also been adapted for domain-specific purposes. Examples include 
       PolyMage \cite{Polymage}, designed for image processing pipelines, and PENCIL \cite{PENCIL}, 
       an approach for constructing parallelizing compilers for DSLs.

    \section{Conclusion}       
    In this paper, we presented a new mathematical framework—Tensor Evolution (TeV).
     TeV extends the theory of Chain of Recurrences (CR) to the evolution of tensor values
      within loops found in machine learning (ML) and high-performance computing (HPC) graphs.
       While CR was originally developed for scalar values and successfully implemented in LLVM and GCC as Scalar Evolution (SCEV), TeV introduces a novel approach by generalizing these concepts to tensors.
    We demonstrated the usefulness of TeV in specific contexts through illustrative code examples.
     The mathematical framework we propose has potential applications in general ML and HPC
      scenarios when further extended and applied creatively. For example, SCEV is widely
       used in optimizations like loop strength reduction (LSR) and loop-exit value computation.
        Similarly, TeV could be leveraged in cases where parts of a tensor evolve predictably,
         enabling the effective application of recurrence lemmas to optimize computations.

\bibliography{reference}{}
\bibliographystyle{unrst}
    
\end{document}